\documentclass{mn2e}
\usepackage{epsf}
\newcommand{\etal}{{et al}\/.}
\newcommand{\pks}{\hbox{PKS\,0521$-$365}}
\newcommand{\chandra}{{\it Chandra}\rm}
\newcommand{\rosat}{{\it ROSAT}\rm}
\begin{document}
\title[\pks]{The X-ray jet and halo of \pks}
\author[M. Birkinshaw et al.]{M. Birkinshaw, D.M. Worrall and M.J. Hardcastle\\
        Department of Physics, University of Bristol, Tyndall Avenue,
        Bristol BS8~1TL}
\date{Accepted \qquad \qquad . Received \qquad \qquad ,
      in original form 4-July-2001}
%
%
\maketitle
\label{firstpage}
\begin{abstract}
\chandra\ ACIS observations of \pks\ find that the X-ray emission of
this BL~Lac object consists of emission from an unresolved core, a
diffuse halo, and a 2~arcsec jet feature coincident with the inner
radio/optical jet. A comparison with a new ATCA 8.6-GHz map also finds
X-ray emission from the bright hotspot south-east of the nucleus. The
jet spectrum, from radio to X-ray, is probably synchrotron emission
from an electron population with a broken power-law energy
distribution, and resembles the spectra seen from the jets of low-power
(FR~I) radio galaxies. The hotspot X-ray flux is consistent with the 
expectations of synchrotron self-Compton emission from a plasma close
to equipartition, as seen in studies of high-power (FR~II) radio
galaxies. While the angular structure of the halo is similar to that
found by an analysis of the \rosat\ HRI image, its brightness is seen
to be lower with \chandra, and the halo is best interpreted as thermal
emission from an atmosphere of similar luminosity to the halos around FR~I
radio galaxies. The X-ray properties of \pks\ are consistent with it
being a foreshortened, beamed, radio galaxy.
\end{abstract}

\section{Introduction}
\label{sec:intro}

It has been suggested (e.g., Antonucci \& Ulvestad 1985; Ulrich 1989;
Urry \& Padovani 1995) that BL~Lac objects are ``unified'' with FR~I
radio galaxies (FRIRGs), so that the only difference between BL Lac
objects and FRIRGs is their orientation relative to the observer's
line of sight. Thus BL~Lac objects are supposed to be viewed from a
direction close to the nuclear jet, whose Doppler-boosted emission
dominates the appearance of the object at radio to $\gamma$-ray
energies. FRIRGs are supposed to be viewed from a direction outside
the nuclear jet, so that the nucleus is faint relative to the outer
radio structure, the optical galaxy, and the unbeamed X-ray emission. 

This idea has been tested through comparisons of BL~Lac and FRIRG
number densities, extended (and hence isotropic) radio luminosities
and morphologies, and the optical luminosities and morphologies of the
underlying galaxies (Morris et al. 1991; Perlman \& Stocke 1993; Pesce,
Falomo \& Treves 1996). Tests based on X-ray properties, which are
intrinsically among the most interesting since strong X-ray emission
is a defining feature of BL Lac objects, have fitted the
distribution of X-ray luminosity to a theoretical function based
on some beaming factor and beam opening angle, using either the total
X-ray luminosity (e.g., Padovani \& Urry 1990) or the core X-ray
luminosity, after any extended X-ray emission has been removed (Canosa
2000; Hardcastle et al. 2001).

Another possible X-ray test of the identification of BL~Lac objects
as beamed FRIRGs can be made based on the association of FRIRGs with
strong X-ray emission from galaxy-to-group scale gas distributions
(Worrall \& Birkinshaw 1994, 2000). Since this emission is thermal and
extended, it must radiate isotropically, and therefore should appear
with similar properties around BL~Lac objects and FRIRGs. A clear
difference between the extended X-ray emissions near BL~Lacs and
FRIRGs would constitute evidence against a strong unification
hypothesis.

This test is difficult because the extended X-ray emission has low
contrast relative to the BL~Lac's unresolved active galactic nucleus
(AGN). Wings on the point response function (PRF) of the X-ray optics
and detector used to observe a BL~Lac are therefore likely to exceed
the brightness of the extended emission. Nevertheless, Hardcastle,
Worrall \& Birkinshaw (1999), using the \rosat\ HRI, reported a halo
of resolved emission around the AGN in the low-redshift ($z = 0.05534$;
Keel 1985) BL~Lac \pks. Although \pks\ had a low X-ray flux during this
observation, the halo was only detectable above the wings on the HRI's
PRF over a small range of angles. The inferred halo luminosity, of 
$L_{\rm X}(0.2 - 1.9 \ \rm keV) \approx 8 \times 10^{35}\ \rm W$, was
far higher than the values $\la 10^{35} \ \rm W$ typical
of the FRIRGs in the B2 sample of Worrall \& Birkinshaw
(1994). Since the core radius of the \pks\ X-ray halo was found to be
only 8~arcsec (12~kpc for $H_0 = 50 \ \rm km \, s^{-1} \, kpc^{-1}$,
$q_0 = 0$, as assumed throughout this paper), the hot gas appeared to be
confined near the host galaxy, rather than being associated with the
group of galaxies in which \pks\ lies, as is more usual for FRIRGs..

At this luminosity the cooling time of gas in the X-ray halo is less
than $3 \times 10^8$~yr, so that a halo of this size and luminosity
would be expected to be participating in a rapid cooling flow, with an
estimated infall rate $> 20 \ \rm M_\odot \, year^{-1}$. The 
\rosat\ HRI data could not measure the temperature of the gas, or say
whether the extended halo had the structure of a cooling flow near the AGN. 

For these reasons, we obtained an imaging observation of \pks\ with
the \chandra\ ACIS-S. We anticipated that this would
provide detailed structural and spectral information on the various
components of the source, including the \rosat-detected halo.

\chandra\ has shown that X-ray jets are relatively common in
low-power radio galaxies (Worrall, Birkinshaw \& Hardcastle
2001). Such jets are best 
interpreted as synchrotron emission from a high-energy continuation of
the same electron spectrum responsible for the radio output. Thus if
BL~Lac objects are simply FRIRGs seen in projection, we might expect to
see the same X-radiation from the jets as is seen in radio
galaxies, with a similar X-ray to radio power ratio. However, this
requires that the observation has enough angular resolution to measure
a foreshortened jet feature close to the Doppler-enhanced X-ray
core. \pks\ is known to exhibit a radio/optical jet on arcsecond
scales (Keel 1986; Macchetto \etal\ 1991) which connects to an inner
VLBI jet on milliarcsec scales (Tingay et al. 1996), and so the arcsec
jet might be resolvable from the bright core by \chandra.

Our observation of \pks, made early in the \chandra\ mission,
successfully detected and studied the halo, and found X-radiation
associated with both the $1.5$~arcsec-scale inner jet of the BL~Lac
and the hotspot of its radio structure. In this paper we describe
these \chandra\ results, and their interpretation in terms of emission
mechanisms and unified schemes for BL~Lac objects.

\section{Observations}
\label{sec:obs}

\subsection{X-ray observations}
\label{sec:xrayobs}

The \chandra\ observation of \pks\ was made on 31~December~1999. The
telescope was pointed so that the source was located near the
aim-point of the back-illuminated CCD chip S3 of the Advanced CCD
Imaging Spectrometer (ACIS). Unfortunately, at this early stage of the
mission the standard offsets placed \pks\ close to the boundary
between two readout nodes of the S3 chip, so that the source moved
between regions of the chip with somewhat different responses, and the
dead region between them, as the pointing direction of the observatory
was dithered by about 20~arcsec.

Since \pks\ was expected to produce a high X-ray count rate, in excess
of 1~count $\rm s^{-1}$, the observation was made with the minimum
permitted ACIS frame time (0.4~sec). To achieve this it was necessary
to observe with a small window on the CCD. This window was 128~pixels
(each of $0.492$~arcsec on a side) wide, and 1024~pixels long, and at
the roll of the satellite during these observations
($340^\circ\llap{.}4$), the long axis of the viewing window lies in
position angle $-70^\circ$

\begin{figure}
 \epsfxsize 8.6cm
 \epsfbox{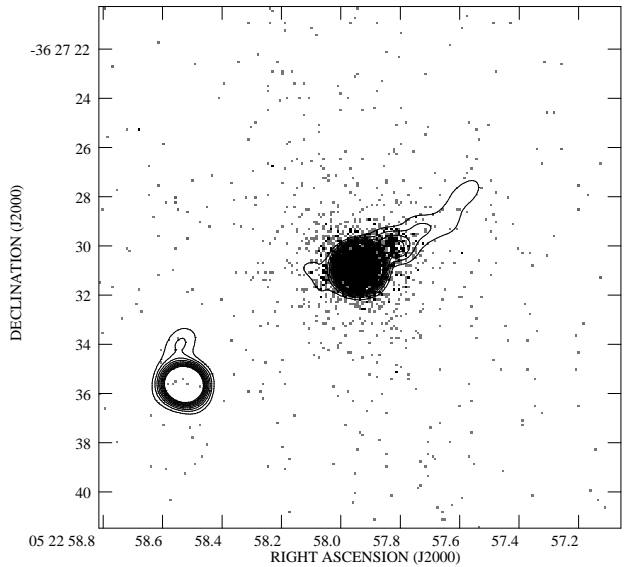}
 \caption{The central part of the subsampled ($0.0984$~arcsec square
  pixels), exposure corrected, $0.3 - 7.0 \ \rm keV$ \chandra\ ACIS-S
  image of \pks. The linear greyscale corresponds to $0 \ \rm to \ 4
  \times 10^{-7} \ photons \, cm^{-2} \, s^{-1} \, pixel^{-1}$. The radio
  image superimposed is our 8.6-GHz map made with the Australia Telescope
  Compact Array with 1~arcsec resolution, contoured at intervals of
  20~mJy from 20~to~200~mJy.}
 \label{fig:rximage}
\end{figure}

Even with the reduced frame time, there is a significant chance that
several photons will be received from the core of \pks\ in a single
CCD pixel within a single integration. This ``pileup'' effect leads to
a distortion of the spectrum of the core and a reduction in the count
rate relative to the true incoming rate. Another consequence of pileup
is that the image of the core differs from the images of fainter,
unresolved, sources, in the sense of appearing somewhat extended and
flat-topped. Finally, because the count rate is high, there is a
significant probability that the CCD will record photons from the
source while the data are being clocked to the readout registers. This
causes faint ``readout streaks'', which appear in position angles
$20^\circ$ and $200^\circ$ from the core of the source for the
$-70^\circ$ roll of these data.

The data were provided to us in a variety of processing states: the
analysis reported here was performed on data processed using pipeline
software version \hbox{\sc R4CU5UPD11.1}, and used software from the
{\sc CIAO} \hbox{v2.0.2}. After screening out periods of high
background, the original exposure of 9122~sec (live time, from an
on-source time of 9870~sec) was reduced to 7776~sec of live time.

An image was made from the events file by restricting the apparent
photon energies to the relatively well-calibrated range $0.3 -
7.0$~keV, and corrections were made for the effective area of the
instrument and the exposure. At the same time, the data were
reprocessed to remove the anti-aliasing pixel randomization included
in the standard pipeline, to produce an image with the best possible
angular resolution, and were resampled onto a pixel scale one fifth
that of the ACIS-S pixels ($0.0984$~arcsec). The resulting image is
shown in Fig.~\ref{fig:rximage}. Clearly visible in this image is the
bright concentration of counts at the core of the AGN, an asymmetric
excess of counts associated with the brightest feature in the radio
jet to the north-west (at J2000 coordinates RA $\rm 05^h 22^m
57^s\llap{.}86$, Dec $-36^\circ 27^\prime
30^{\prime\prime}\llap{.}1$), and a radial gradient in the count
density from the AGN. An excess of counts is also be associated
with the bright hotspot of the radio structure, to the south-west of
the source.

\begin{figure}
 \epsfxsize 8.6cm
 \epsfbox{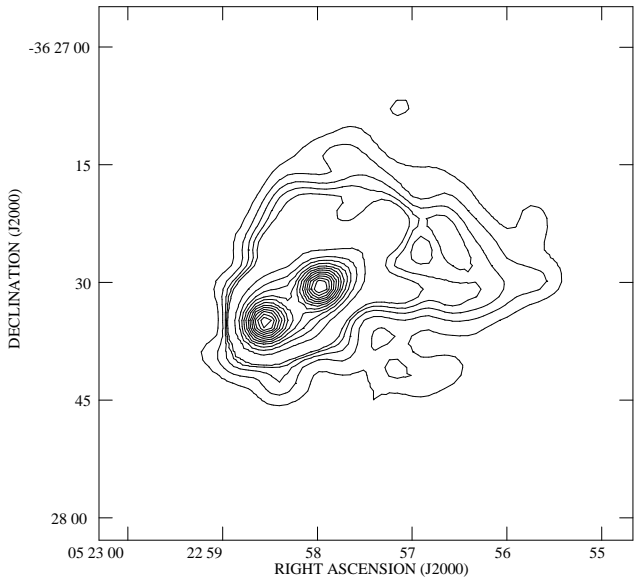}
 \caption{A 2.5-GHz radio map of \pks\ made with the Australia
  Telescope Compact Array, with 4~arcsec resolution, and contoured at
  intervals of 20~mJy from 20~to~100~mJy, then every 200~mJy. Bright
  compact components are associated with the core and the hotspot
  to its south-east. A halo of lower surface brightness emission
  surrounds these components.}
 \label{fig:simage}
\end{figure}

Spectral analyses were performed using the {\sc CIAO} {\sc SHERPA}
package, after calculating the response matrices in the standard
fashion. Although the matrices used for fitting were appropriate for
only one of the readout nodes of the ACIS-S3 chip (the node on which
most of the counts lie), tests using matrices calculated for the other
readout node showed no significant differences from the results
presented here. Fits were performed only in the relatively
well-calibrated energy range $0.3 - 7.0 \ \rm keV$, and only after the
counts had been binned such that there were 20~or more counts in each
energy bin. 

\begin{figure}
 \epsfxsize 8.6cm
 \epsfbox{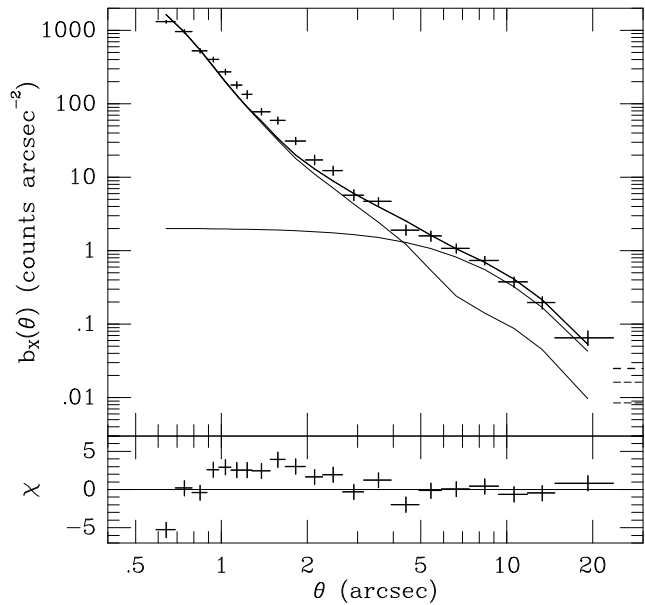}
 \caption{The radial profile of \pks\ extracted from the subsampled
  image (Fig.~\ref{fig:rximage}), shown as datapoints. Superimposed is
  the best-fitting composite model made up of a point source (using a
  point response function model for the \chandra\ ACIS-S of the form
  discussed by Worrall et al. 2001) and a beta-model with $\beta = 0.90$
  and the best-fitting core radius of $9.3$~arcsec. The fit in the inner
  2~arcsec is poor, most likely because of the effects of the readout
  streak and pileup, but the outer region is well described by a
  superposed beta model and the wings of the point source.}  
 \label{fig:radprofile}
\end{figure}

A radial profile centred on the core of \pks\ was extracted from the
subsampled, $0.3 - 7.0$~keV image, and analysed in the manner
described in Worrall \etal\ (2001), using an
averaged point response function based on the {\sc CIAO} PRF library
and the spectra of several similar sources. This radial
profile (shown in Fig.~\ref{fig:radprofile}) is expected to be a poor
representation of the true source structure near the core because of
pileup: fits to the profile therefore ignored the central 1.2~arcsec
diameter circle. In addition, the profile was constructed after
omitting the jet region and the readout streaks. The final profile
contains only about a third of the detected counts within a radius of
23.5~arcsec.

\subsection{Radio observations}
\label{sec:radioobs}

To complement these X-ray data, we mapped \pks\ using the Australia
Telescope Compact Array (ATCA) in its 6D configuration on 4~and
5~April~2000, giving baselines from 77~to 5878~m. Data were obtained
at 1.384, 2.496, 4.800, and 8.640~GHz with 128~MHz bandwidth
subdivided into 32~channels (of which 16~are independent and only
13~are useful). 10-second integrations were used at all
bands. Observations of \pks\ were calibrated for amplitude against the
standard southern calibrator PKS~1934$-$638, and PKS~0537-441 was used
as the local phase calibrator. 

All six antennas of the ATCA were available for most of the
observation, although difficulties with the correlator caused some 
$uv$ coverage to be lost, and some of the data were taken in bad
weather, so that the phase stability was sometimes poor. Phase
calibrations were made every 15 minutes to compensate for the unstable
conditions, but after basic calibration and data-editing using 
{\tt MIRIAD}, a lengthy process of self-calibration 
(in {\tt AIPS}) was needed to correct the phases and improve the
images. Only a small fraction of the data were lost because of bad
amplitudes, interference, or phase jumps. Residual phase errors limit
the dynamic range of the final image to about $10^4$.

\begin{figure}
 \epsfxsize 8.6cm
 \epsfbox{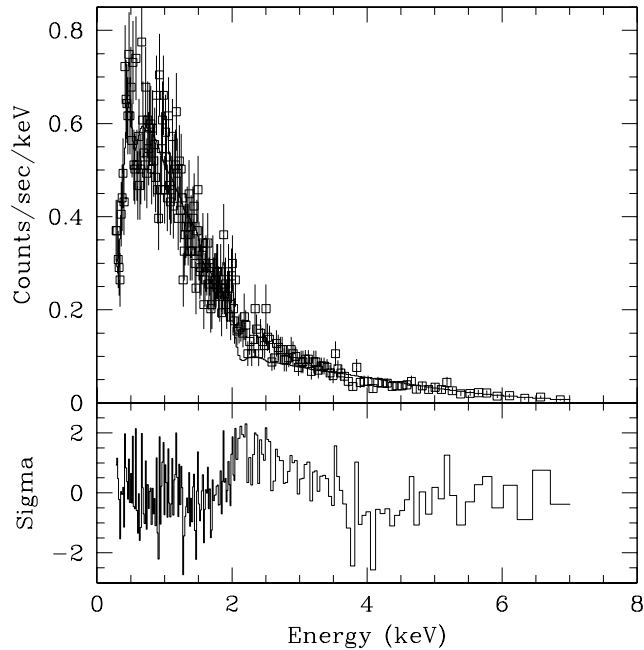}
 \caption{The spectrum of the core of \pks\ in energy range 
  $0.3 - 7.0$~keV, for 
  which the calibration is expected to be adequate, with the best-fitting
  absorbed power-law spectrum (upper panel). The lower panel shows the 
  residuals from this fit. Note the systematic misfit near 2~keV, which
  arises from pileup of photons near 1~keV, where the count rate
  peaks. The fit is of adequate quality ($\chi^2 = 215$ with 199~degrees
  of freedom), but biased by pileup.
 }
 \label{fig:xcorespec}
\end{figure}

Fig.~\ref{fig:rximage} shows contours of the 8.6-GHz, 1-arcsec
resolution radio map of \pks\ superimposed on an image of the 
0.3-7.0~keV X-ray counts derived from our \chandra\ ACIS-S data.
The radio map shows an unresolved component at the core of \pks, a
clear jet to the north-west dominated by a knot about 1.5~arcsec from the
core, and a compact hotspot to the south-east. A Gaussian fit to the
hotspot measures a FWHM of 0.4~arcsec. This result will be important in
Sec.~\ref{sec:hotspot}. 

Much low-surface-brightness emission does not appear on
Fig.~\ref{fig:rximage} at these contour levels, but is revealed on 
a lower-resolution map of \pks, made with our ATCA 2.5-GHz data, and
shown in Fig.~\ref{fig:simage}. The diffuse, extended, radio emission
associated with the source is seen to surround the radio core and the
hotspot to the south-east. Only a slight extension of the radio core
to the north-west betrays the presence of the radio jet.

\begin{figure}
\epsfxsize 8.6cm
\epsfbox{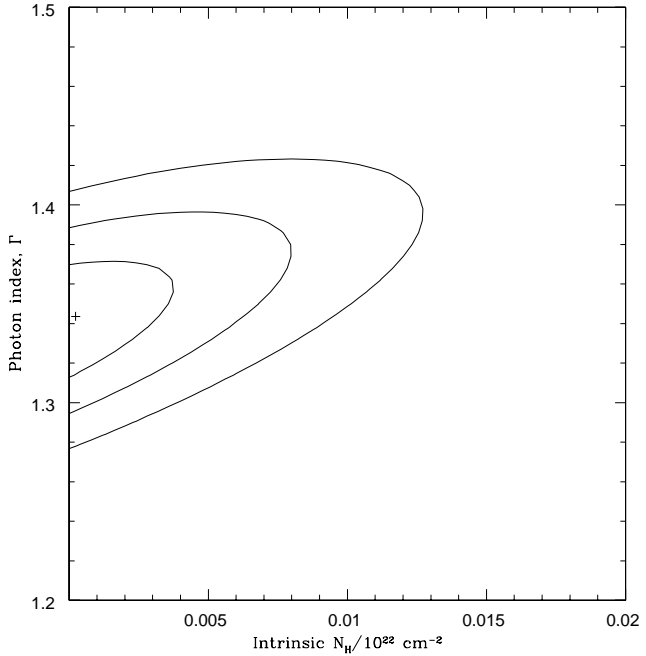}
\caption{$\chi^2$ contours for fits to the spectrum of \pks\ 
as a function of the photon index of the fitted spectrum,
$\Gamma$, and the intrinsic absorbing column at the core of 
\pks, $N_{\rm H}$. Contours are drawn at $1$, $2$, and
$3\sigma$ uncertainty offsets from the best fit. No excess
$N_{\rm H}$ at \pks\ is required by these fits.}
\label{fig:corepars}
\end{figure}

The distorted radio morphology of \pks\ evident in these maps
is characteristic of radio-selected BL~Lac objects, but \pks\ is 
somewhat unusual in showing the bright jet typical of FR~I radio
galaxies to the north-west combined with a hotspot, typical of FR~II
radio galaxies, to the south-east. FR~II-like and FR~I-like
extended radio morphologies are about equally common in radio-selected
BL~Lac objects (Rector \& Stocke 2001). It is not implausible,
therefore,  that a BL~Lac should show a mixed-morphology radio
structure, which may be related to the structures of radio galaxies
with luminosities near the FR~I/FR~II break (Gopal Krishna \& Wiita
2000).

\section{Analysis}
\label{sec:anal}

\subsection{The core}
\label{sec:analcore}

The X-ray image of \pks\ in Fig.~\ref{fig:rximage} is dominated by a
strong core component, responsible for $1.01 \pm 0.02 \ \rm count \,
s^{-1}$ in a 2-arcsec diameter circle. At this count rate the
core is strongly piled up, and individual counts frequently correspond
to the combined energies of two or three photons that arrived during a
single frame. Nevertheless a spectrum was extracted and a spectral fit 
was attempted, using counts from a region 2.2~arcsec in
diameter about the centre of the core and the standard
choice of event grades (ASCA grades 0, 2, 3, 4, and 6). The effects of
pileup in the resulting spectrum (Fig.~\ref{fig:xcorespec}) are clear
from the shallowness of the Ir edge (at about 2~keV) introduced by the
mirror coatings. This edge feature is suppressed by the pileup of
the high-count 1~keV photons, which correspondingly are depleted in
the observed spectrum.

No {\sc CIAO} task exists to correct for pileup at present, and
therefore fits to the core spectrum will be subject to a systematic
error in the sense of underestimating the core flux density at low
energies and underestimating the spectral index of the core, $\alpha$
(defined in the sense that the flux density $S_\nu \propto
\nu^{-\alpha}$). In addition, this early observation was pointed so
that the core lies on the junction of two readout nodes of the ACIS-S3
chip. In such a location, it is difficult accurately to estimate the
response matrix appropriate for the analysis of the spectrum, and a
further systematic error is possible. 

Nevertheless, some indication of the properties of the core
X-radiation can be deduced from the X-ray spectrum. A fair fit
to the data is obtained by fitting to an absorbed power-law spectrum
(Fig.~\ref{fig:xcorespec}). The best-fitting spectrum has a spectral 
index $\alpha = 0.34 \pm 0.03$ (corresponding to a photon index $\Gamma =
\alpha + 1 = 1.34$), consistent with the spectral index found by Garilli \&
Maccagni (1990) using EXOSAT, but flatter than the value
$\alpha = 0.93 \pm 0.03$ reported by Hardcastle \etal\ (1999) based on
the \rosat\ PSPC data. The effects of pileup and inadequate
calibration of the response matrix may cause a systematic
error of about $0.1$ on the measured power-law index. The corresponding
best-fit 1~keV photon flux density, $(1.03 \pm 0.02) \times 10^{-3} \
\rm photons \, cm^{-2} \, keV^{-1} \, s^{-1}$ ($0.68 \pm 0.01 \ \rm
mJy$), is likely to be a serious underestimate (and, indeed, is a
factor $\approx 3$ below the flux density measured by the 
\rosat\ PSPC). If account is taken of the 
Galactic column towards \pks\ ($N_{\rm H} = 3.37 \times 10^{20}
\ \rm cm^{-2}$; Elvis, Wilkes \& Lockman 1989), 
no absorbing column associated with the AGN is required in this fit
(Fig.~\ref{fig:corepars}): the $2\sigma$ upper limit is
$N_{\rm H} < 8 \times 10^{19} \ \rm cm^{-2}$. Pileup and calibration 
uncertainties affect the shape of the core spectrum below the peak,
but an intrinsic absorbing column significantly larger than this limit
would strongly affect the cut-off energy in Fig.~\ref{fig:xcorespec},
so that we believe that the limit on $N_{\rm H}$ quoted above is
reliable, and indicates that we are observing the core emission
free from an absorbing column as large as found towards the cores of
FRIRGs. 

No correction for background was needed in making these spectral fits:
a nearby background region with the same area as the source extraction
region contained only 4~counts (about 3~of which arise from the
wings of the response from the core), compared with the 7900~counts in
the spectrum shown in Fig.~\ref{fig:xcorespec}.

\begin{figure}
\epsfxsize 8.6cm
\epsfbox{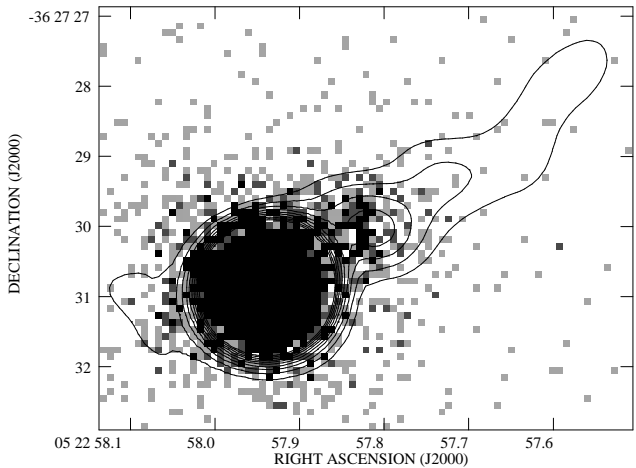}
\caption{An expanded view of the core and jet region, with the same
grey scale and contours as in Fig.~\ref{fig:rximage}. The X-ray
structure associated with the bright knot at the base of the radio jet
is seen to lie roughly perpendicular to the radio axis.}
\label{fig:inner}
\end{figure}

\subsection{The jet}
\label{sec:analjet}

A close examination of the structure of the X-ray emission from the
jet knot (knot A in the notation of Scarpa et al. 1999) appears to
show an elongated structure with its long axis lying in
position angle $(7 \pm 2)^\circ$. This is roughly perpendicular to the
local ridge-line of the radio jet (see Fig.~\ref{fig:inner}), and 
differs significantly from the position angles of the chip
pixelization. The elongation may therefore be real despite appearing
exceptionally narrow, and potentially being subject to dither effects
and aliasing in the ACIS pixel assignments. Similar narrow,
jet-crossing, features have been seen in other radio jets (e.g.,
3C\,66B; Hardcastle, Birkinshaw \& Worrall 2001b), and tend to lie
just core-wards of bright jet radio knots. 

The emission from this X-ray knot is clearly contaminated by counts
from the wings of the response to the AGN. 334 counts at energy $< 10
\ \rm keV$ are found from knot A, where the counts were extracted from
a circular region with diameter 2.2~arcsec centred on the knot.
A similar extraction region on the opposite side of the core was used
for background, since it was expected to be equally contaminated by
counts from the wings of the core. This background region contains
92~counts. Neither the jet knot nor the background region show
count rates high enough that pileup is a problem. Thus about a quarter
of the counts from the region centred on the jet knot can be
attributed to contamination from the core, or emission from the thermal halo.
After correcting for this background, we find that the knot is
detected by \chandra\ at $0.031 \pm 0.003 \ \rm \ count \, s^{-1}$ in
$0.3 - 7.0$~keV.

\begin{figure}
\epsfxsize 8.6cm
\epsfbox{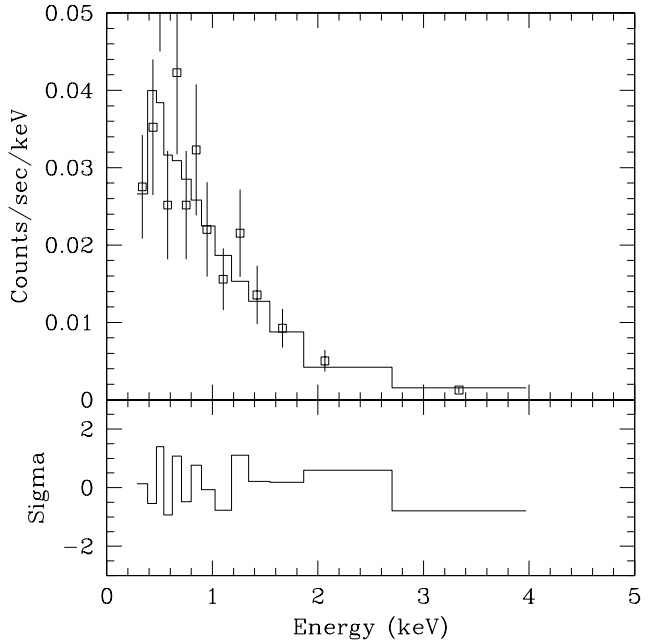}
\caption{The spectrum of a 2.2-arcsec diameter region centred on the
 X-ray knot (upper panel) and the residuals from the fit (lower
 panel). The best fitting power-law model for the knot spectrum is
 shown as the solid line, and corresponds to energy index $\alpha =
 1.41 \pm 0.57$ (photon index $\Gamma = 2.41$), 1-keV flux density $15
 \pm 5 \ \rm nJy$, and an absorbing column of $(1.7 \pm 4.7) \times
 10^{20} \ \rm cm^{-2}$ at the redshift of \pks. The background from
 the wings of the point response of the core has not been subtracted in
 this spectrum, and is simultaneously fitted. A good description of the
 background is given by a power law with energy index $\alpha = 1.58
 \pm 0.20$ and a 1-keV flux density about 45~per cent that of the jet
 knot. The quality of the fit is good: $\chi^2 = 13.9$ with 23~degrees
 of freedom in background and source combined.
}
\label{fig:xjetspec}
\end{figure}

The spectrum extracted from the jet region {\it including
background from the core and thermal halo} is shown in
Fig.~\ref{fig:xjetspec}. Sufficient counts exist for meaningful
fits only in the energy range $0.3 - 4.4$~keV. The fits were
performed using {\sc SHERPA}'s ability to fit source and background
simultaneously. The jet emission was modelled as a 
power-law, absorbed by the Galactic column and a possible excess at
the redshift of \pks. The background was modelled as a simple
power-law (found to have a similar photon index to that of the jet
knot; Fig.~\ref{fig:xjetspec}), which was expected to be a good
representation of the superposition of the peripheral counts from the
core of the the AGN and the thermal halo.

\begin{figure}
\epsfxsize 8.6cm
\epsfbox{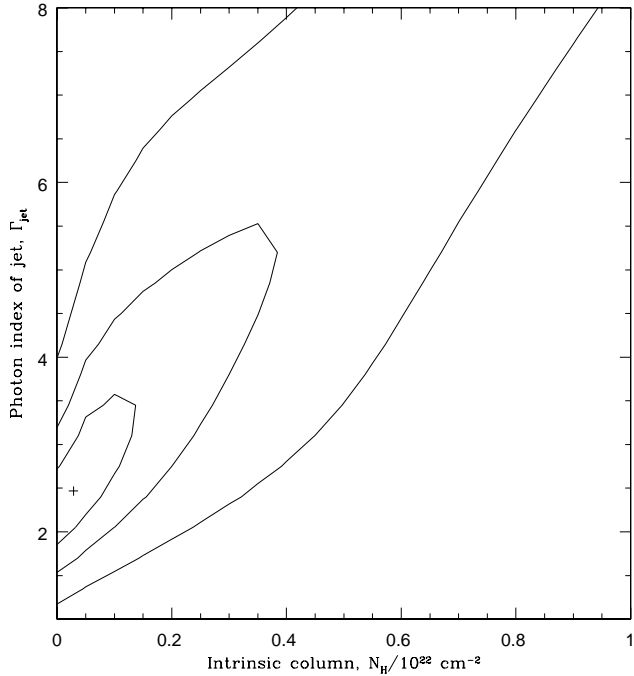}
\caption{$\chi^2$ contours for fits to the spectrum of the jet of
 \pks\ as a function of the photon index of the fitted
 spectrum, $\Gamma_{\rm jet}$, and the intrinsic absorbing column in
 front of the jet, $N_{\rm H}$. Contours are drawn at $1$, $2$, and
 $3\sigma$ uncertainty offsets from the best fit, which lies at $\chi^2
 = 13.9$, $\Gamma = 2.41$, $N_{\rm H} = 1.7 \times 10^{20} \ \rm
 cm^{-2}$. The best fit with no such absorption has
 $\chi^2 = 14.0$ (with one additional degree of freedom) and photon
 index $\Gamma = 2.25$.}
\label{fig:xjetpars}
\end{figure}

The best fit (shown in Fig.~\ref{fig:xjetspec}) involves an
intrinsic absorbing column of $1.7 \times 10^{20} \ \rm cm^{-2}$, but
this is not significant (Fig.~\ref{fig:xjetpars}). If the additional
column is set to zero, as suggested by the fit to the core spectrum,
the jet energy spectral index changes from $1.41 \pm 0.57$ to $1.25
\pm 0.30$ ($1\sigma$ errors with one interesting parameter). The
significant change in the error on the spectral index arises from 
the strong correlation between the fitted power-law index and the
poorly-constrained absorbing column (Fig.~\ref{fig:xjetpars}). The
value of $\chi^2$ at the best fit ($13.9$, with 23~degrees of freedom)
is only slightly less than the value with no absorbing column ($\chi^2
= 14.0$, with 24~degrees of freedom). None of the other fitted
parameters vary by more than 10~per cent of their errors if it
is assumed that there is no intrinsic absorbing column at \pks.

\begin{figure}
\epsfxsize 8.6cm
\epsfbox{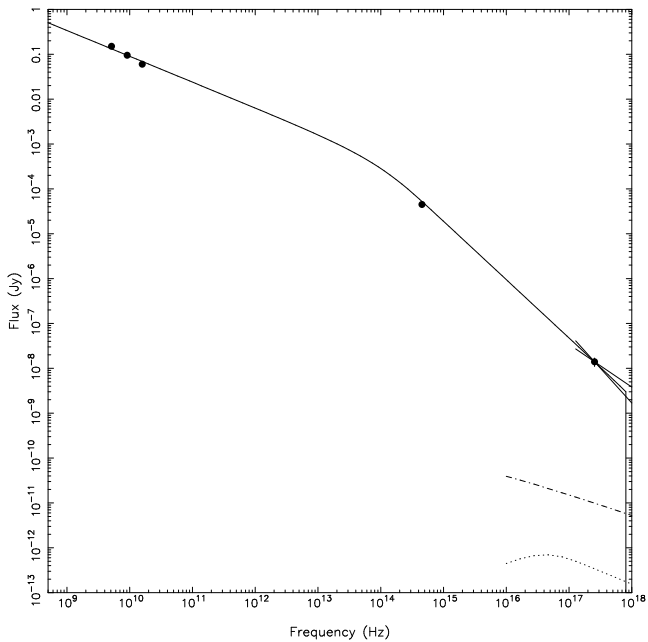}
\caption{A synchrotron model roughly matching the radio,
 optical, and X-ray flux densities of the bright knot in the jet of
 \pks as given in Table~\ref{tab:knotsed}. The R-band and 15-GHz data
 are taken from Scarpa et al. (1999), 
 the other radio flux densities are from our ATCA and archival VLA
 data, and the X-ray spectrum is shown with the 1-keV flux density and
 the permitted range of spectral indices assuming that the intrinsic
 absorbing column is zero. The model requires a break in the infra-red
 of about $\Delta \alpha = 0.7$, corresponding to an electron energy of
 about $200 \ \rm GeV$ in the equipartition field. Neither the
 synchrotron self-Compton radiation (dash-dotted line) nor the
 inverse-Compton radiation produced by scattering the microwave
 background radiation (dotted line) makes a significant contribution to
 the observed X-ray flux of the jet.
}
\label{fig:jetsed}
\end{figure}

The 1-keV flux density of the jet knot, from the best fit without
intrinsic absorption, is $14 \pm 3 \ \rm nJy$, which corresponds to an
$0.2 - 2.4$~keV luminosity of $1.0 \times 10^{35} \ \rm W$ if the
emission is isotropic. The spectral energy distribution of this knot
is shown in Fig.~\ref{fig:jetsed}, where flux densities
(given in Table~\ref{tab:knotsed}) are for a nominal 2.2-arcsec
diameter aperture centred on the X-ray knot. The solid
line drawn on Fig.~\ref{fig:jetsed} is calculated as the synchrotron
emission from a population of electrons with a broken power-law energy 
distribution. 

On Fig.~\ref{fig:jetsed} the radio spectrum appears steeper by about
$0.2$ than the model, which has a radio spectral index of $0.58$
(corresponding to an electron energy index $p = 2.16$), probably
because of a contribution to the lower-frequency flux densities from
steeper-spectrum emission from the halo. The X-ray spectrum is
significantly steeper, with $\alpha \approx 1.25$. This extrapolates
back to a good match to the R-band optical flux measured by Scarpa et
al. (1999), and implies that the 
spectrum must break by $\Delta\alpha \approx 0.7$ in the
infra-red. For the equipartition magnetic field strength of $16 \ \rm
nT$ (estimated assuming that the knot is moving non-relativistically,
has an angular diameter of $0.5 \ \rm arcsec$, is completely filled by
magnetic fields and relativistic electrons of energies $> 150 \ \rm
MeV$, and that there is a negligible contribution to the total energy
from protons), this break lies at an electron energy of about $200 \
\rm GeV$.

\begin{table}
 \caption{Jet knot flux density measurements}
 \label{tab:knotsed}
 \begin{tabular}{@{}cll}
  Frequency (Hz)                 & Flux density
                                 & Reference \\
  $4.9 \times 10^{9\phantom{1}}$ & $          151 \pm 5 \ \rm mJy   $ 
                                 & this work \\
  $8.6 \times 10^{9\phantom{1}}$ & $\phantom{1}95 \pm 5 \ \rm mJy   $ 
                                 & this work \\
  $1.5 \times 10^{10}          $ & $\phantom{1}60 \pm 5 \ \rm mJy   $ 
                                 & Scarpa et al. (1999) \\
  $4.3 \times 10^{14}          $ & $\phantom{1}45 \pm 6 \ \rm \mu Jy$
                                 & Scarpa et al. (1999) \\
  $2.4 \times 10^{17}          $ & $\phantom{1}14 \pm 3 \ \rm nJy   $
                                 & this work \\
 \end{tabular}

\end{table}

The amplitude $\Delta\alpha \approx 0.7$ of the break in the spectrum
corresponds to a change in electron energy power-law index $\Delta p
\approx 1.4$. Such a large break is unexpected from simple synchrotron
ageing arguments, and suggests that there is a change in efficiency of
the electron acceleration process at about $200 \ \rm GeV$. In the
shape of the inferred electron energy distribution and the value of
the equipartition magnetic field, this knot is very similar to that in
3C~66B.

Also plotted on Fig.~\ref{fig:jetsed} are the expected levels of
inverse-Compton emission from the electrons scattering the microwave
background radiation and the jet's own synchrotron emission, assuming
that the population of electrons in the jet is given by equipartition
arguments. The turndown in the predicted inverse-Compton emission from
scattering the microwave background radiation occurs because of the
assumed lower energy limit of $150 \ \rm MeV$ for the electron
population. Neither scattering process can be responsible for as much
as $10^{-3}$ of the X-ray emission seen under the assumption that the
jet is non-relativistic.

If the jet speed is highly relativistic, then the level of
inverse-Compton emission from scattering the microwave background is
higher than modelled in Fig.~\ref{fig:jetsed}, with a Lorentz factor
$\approx 30$ being sufficient to produce the observed X-ray
brightness. However, even in this case the predicted spectrum would be
too flat to match the steep X-ray spectrum measured by \chandra\ unless
there is a substantial population of low-energy electrons (which
must emit radio synchrotron radiation only at frequencies below $5 \
\rm GHz$). An additional photon source, for example beamed emission
from the nucleus, can also add extra inverse-Compton X-ray emission,
but again the predicted spectrum would be a poor match to the data
unless a low-energy electron population is present.

Thus, since the X-ray spectrum produced by inverse-Compton X-ray
emission would have $\alpha \approx 0.6$ (the same spectral index
as the source photons, which are upscattered principally from the
radio and mm-wave bands), while the observed spectrum has $\alpha =
1.25 \pm 0.30$, we conclude that the jet X-ray emission in \pks\
arises from synchrotron radiation, as it does in FRIRGs (Hardcastle et
al. 2001b; Worrall et al. 2001). The appearance of the spectral energy
distribution from \pks\ (Fig.~\ref{fig:jetsed}) also strongly resembles
that seen in FRIRGs.

\subsection{The halo}
\label{sec:analhalo}

Hardcastle et al.'s (1999) analysis of the \rosat\ HRI data implied a
bright X-ray halo about \pks. At the best-fit halo brightness, and
assuming that the temperature of gas responsible for the halo is
$\approx 1 \ \rm keV$ we would have expected \chandra\ to detect about
10~times as many counts from the halo in Fig.~\ref{fig:rximage} as
were found in the \rosat\ image. However, the radial profile
(Fig.~\ref{fig:radprofile}) shows a much fainter level of emission. At
10~arcsec from the core, where the excess above the PRF wings is
strongest, the halo contributes about $5 \times 10^{-5} \ \rm count \,
arcsec^{-2} \, s^{-1}$. The earlier fits to the \rosat\ HRI data showed
a best-fit brightness at 10~arcsec of about $2 \times 10^{-5} \ \rm
HRI \ count \, arcsec^{-2} \, s^{-1}$, which would suggest an ACIS
brightness of $2 \times 10^{-4} \ \rm count \, arcsec^{-2} \, s^{-1}$,
four times brighter than the halo that appears in
Fig.~\ref{fig:radprofile}.

\begin{figure}
\epsfxsize 8.6cm
\epsfbox{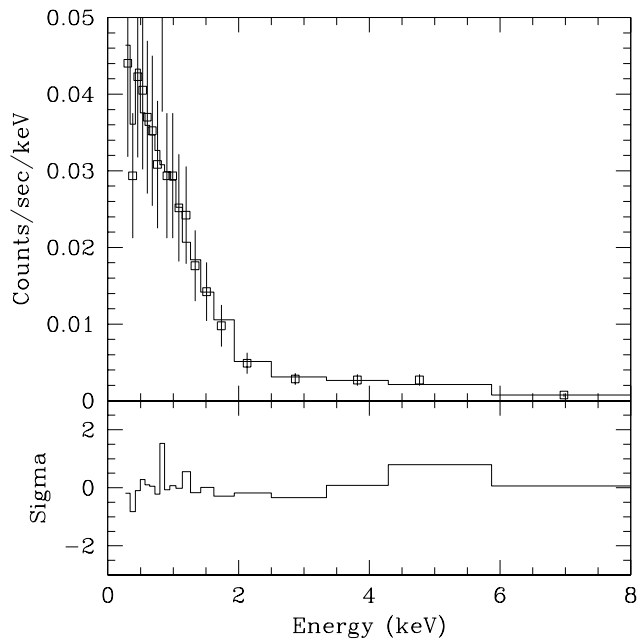}
\caption{The spectrum of the X-ray halo (upper panel) and the
 residuals from the fit (lower panel). The best fitting Raymond-Smith
 thermal model is shown as the solid line, and corresponds to a model
 with $k_{\rm B} T = 1.58 \ \rm keV$, abundance $0.06$ solar, and no
 absorbing column above the Galactic level. Background emission is
 also present in this spectrum, and contributes about half the total
 counts (especially above 3~keV). This background is simultaneously
 fitted, and can be well described by two power laws, with 
 photon indices $\Gamma = 2.46$ and $-0.13$. The quality of the fit is
 good ($\chi^2 = 13.3$ with 35~degrees of freedom in background and
 source). 
}
\label{fig:xhalospec}
\end{figure}

However, the error on the brightness of the halo derived from the HRI
image analysis was considerable, since excess counts above those
expected for a bright point source were seen only in the angular range
$10 - 30$~arcsec, with the wings dominating further out and the core
response dominating close in. If a slightly altered HRI point response
function (based on more recent calibrations than were available at the
time of the Hardcastle et al. analysis) is used, the best fit changes
dramatically, from $\beta = 0.90$ to $0.66$, core radius
$8$ to $3.6$~arcsec, and halo brightness at 10~arcsec of
$2 \times 10^{-5}$ to $5 \times 10^{-5} \ \rm HRI \, count \,
arcsec^{-2} \, s^{-1}$. While these results are even less consistent
with the \chandra\ data, the strong change in the fit based
on a small variation in the HRI calibration indicates that large
systematic errors are possible in the HRI-derived halo
parameters. Furthermore, when account is taken of the range of models
allowed by the HRI data, the error on the 10-arcsec count rate is
about 50~per cent.

If the radial profile in Fig.~\ref{fig:radprofile} is 
fitted using a standard point response for the core, and the central
region is excluded because of the misfit anticipated because of
pileup, it is found that the best-fitting $\beta$ model halo has
structural parameters $\beta = 0.90$ and core radius $9.3$~arcsec, 
strikingly similar to the values found by Hardcastle et al.
(1999). Adopting the new \chandra\ normalization of the
halo, however, has the effect of reducing the halo brightness by about
an order of magnitude from the value reported based on the \rosat\ data:
the implied total X-ray emission from the halo of about 
$0.065 \ \rm count \, s^{-1}$ in $0.3 - 7.0 \ \rm keV$, corresponds to
a bolometric X-ray luminosity $L_{\rm X} \approx 3 \times 10^{35}\ \rm
W$ (using a temperature of $1.5 \ \rm keV$ for the halo gas), or $2
\times 10^{35} \ \rm W$ in the $0.2-1.9 \ \rm keV$ band, about
a factor $4$ fainter than was measured by \rosat\ (Hardcastle et
al. 1999). 

\begin{figure}
\epsfxsize 8.6cm
\epsfbox{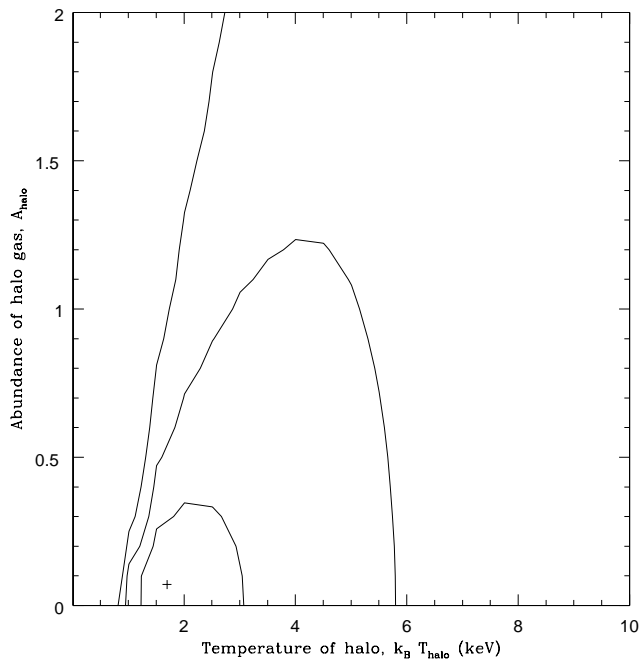}
\caption{$\chi^2$ contours for fits to the spectrum of the halo of
\pks\ as a function of the temperature of the fitted
spectrum, $k_{\rm B} T_{\rm halo}$, and the metal abundance of the
gas, $A_{\rm halo}$. Contours are drawn at $1$, $2$, and
$3\sigma$ uncertainty offsets from the best fit.}
\label{fig:xhaloconf}
\end{figure}

A spectral analysis of the halo was attempted by extracting the counts
from a region between 5 and 20~arcsec from the core, excluding the
readout streaks, and background from 20 to 27~arcsec from the
core. The halo extraction region contains 554 counts, while the
background region contains 212 counts of which about 27 (based on the
best-fitting structural model) arise from the outer 
halo. Roughly 40 counts in the halo region arise from the wings of the
response to the AGN. This complicated situation was dealt with by
making a simultaneous fit to the halo and background spectra, with the
background represented by two power-law spectra: a flat component
that models the particle background, and a steep component that models
the soft X-ray background. This representation of
the background is consistent with the spectra seen in other regions of
the S3~chip. A good fit (Fig.~\ref{fig:xhalospec}) is
found with the halo emission arising from a (Raymond-Smith) thermal
plasma with temperature $k_{\rm B} T = 1.6^{+1.5}_{-0.4} \ \rm keV$,
abundance $0.05^{+0.25}_{-0.05}$~solar ($1\sigma$ errors with
2~interesting parameters, see Fig.~\ref{fig:xhaloconf}), and 1-keV
flux density $130 \pm 20 \ \rm nJy$.  No additional absorbing column
is needed. A thermal spectrum for the halo, with $\chi^2 = 13.3$
and 35~degrees of freedom, is superior to a power-law fit ($\chi^2 =
16.5$ with 36~degrees of freedom). Adding a power-law component to the
halo model, to correct for counts from the wings of the AGN's PRF, has
no significant effect on the goodness of fit or the best-fit
parameters.

The halo temperature and luminosity are similar to the values seen
from the atmospheres of FRIRGs and lie close to the correlation
between these quantities described by Worrall \& Birkinshaw
(2000). Thus the properties of the halo of \pks\ are consistent with
the unification hypothesis, that BL~Lac sources and FRIRGs are from
the same population seen at different orientations. 

\subsection{The hotspot}
\label{sec:hotspot}

\begin{figure}
 \epsfxsize 8.6cm
 \epsfbox{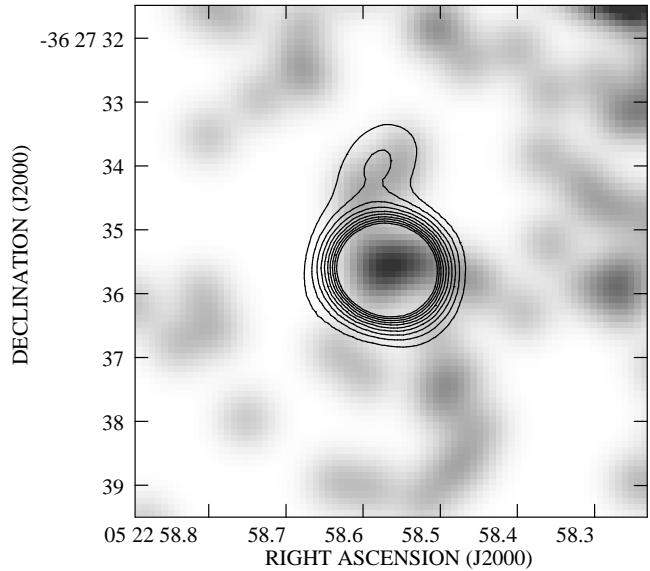}
 \caption{The hotspot of \pks\ from our 8.6-GHz ATCA image, with
  the same contours as Fig.~\ref{fig:rximage}, superimposed on the
  0.3-7.0~keV \chandra\ image convolved to a resolution of
  0.75~arcsec. Note the close alignment of the radio hotspot with a
  relatively bright X-ray feature, which we believe to arise from 
  inverse-Compton scattering of the synchrotron radiation by a
  near-equipartition plasma in the hotspot.} 
 \label{fig:hotspot_image}
\end{figure}

An excess of counts is seen within the 20~mJy contour of the radio map
of Fig.~\ref{fig:rximage}. This 
is better represented in the convolved image of
Fig.~\ref{fig:hotspot_image}, where a clear X-ray peak is
associated with the radio hotspot. We find 5~counts within
1~arcsec of the centre of the hotspot, while the local background
implies that only 1~count would be expected by chance. This
corresponds to a $3\sigma$ significant detection of the hotspot,
although its flux, of about $5 \times 10^{-4} \ \rm count \, s^{-1}$ in
$0.3 - 7.0 \ \rm keV$, is poorly determined. 

The corresponding 1~keV flux density of the hotspot is about 0.4~nJy.
If we model it as an 0.4-arcsec diameter sphere (consistent with its
extension on the 8.6-GHz ATCA map, Fig.~\ref{fig:rximage}), and
assume it to be completely filled with a uniform magnetic field and a
relativistic electron-proton plasma whose energetics are dominated by
the electrons, then assuming equipartition and electron energy limits
$600 \ \rm MeV - 400 \ GeV$, the predicted synchrotron self-Compton
X-ray flux density from the hotspot is about 0.2~nJy.

While the significance of the detection is low, the reasonable
agreement between the observed and predicted X-ray count rates
from the hotspot is consistent with the hotspot being close to
equipartition between its population of relativistic electrons and
magnetic energy. Inhomogeneities in the structure of the hotspot, the
presence of electrons with different energy distributions, or small
departures from equipartition are sufficient to match the observed
X-ray flux density more closely. Similar results have now been seen in
a number of other radio hotspots (e.g., 3C~123; Hardcastle,
Birkinshaw \& Worrall 2001a).

\section{Conclusions}
\label{sec:conc}

Our observation of \pks\ has confirmed the existence of an X-ray
emitting halo around the AGN, but at a lower luminosity than suggested
by the \rosat\ HRI data. The brightness of the halo, as seen with
\chandra\ is consistent with the halo brightnesses seen around FRIRGs,
supporting the idea that this BL~Lac is an FRIRG seen in a favourable
orientation, and that BL~Lac objects do not have unusual X-ray
atmospheres.

The detection of an X-ray bright region near the radio knot in the
jet, though initially surprising, is again consistent with recent
\chandra\ results on FRIRGs. Indeed, the spectral energy
distribution of emission from the knot strongly resembles that from
the inner jet region of 3C~66B (Hardcastle \etal\ 2001b) or M~87
(B\"ohringer \etal\ 2001), and can be fitted well as 
synchrotron emission with a broken power-law spectrum. However, the
break $\Delta \alpha = 0.7$ that we infer in the spectrum is larger
than the value $\Delta \alpha = 0.5$ that would occur in
a continuous-injection model for the relativistic electrons, 
and is also inconsistent with a single-injection aged spectrum, which
would have an exponential cut-off above some maximum energy. The
flattened appearance of the knot X-rays, their slight offset from the
radio peak, and the analogy with FRIRGs, suggests that we are seeing
evidence of energy-dependent electron acceleration.

There is evidence that X-rays are being detected from the
hotspot of \pks\ at a level consistent with equipartition of
relativistic electron and magnetic field energy densities in this
structure. This is consistent with recent results for the
hotspots in many FR~II radio galaxies.

Further study of the intriguing emission structure in the radio jet
demands better optical, near-IR, and sub-mm data to constrain the
shape of the spectral energy distribution and confirm our inference
about the location of the break, but could also benefit from a longer
X-ray exposure to acquire better statistics on the shape of the
emission region in the X-ray. A longer X-ray exposure would also
check our tentative detection of the hotspot. 

Clearly it would be of great interest to discover whether the results
seen for \pks, which suggest that it is a beamed radio galaxy as
required by unification schemes, are common to a sample of BL~Lac
objects rather than a peculiarity of this single source. \pks\ is in
some ways an unusual BL~Lac. Its radio structure displays
characteristics of both FR~I and FR~II sources oriented close to the
line of sight. Although radio-selected (and hence a Radio BL~Lac), it
has been described as a flat-spectrum radio quasar on the basis that
it has a spectral energy distribution peaking at high energies (like
the X-ray selected BL~Lac objects: more strictly, the high-energy
peaked class rather than the low-energy peaked class) and a rather
steep X-ray spectrum (Donato \etal\ 2001). Furthermore, its spectrum
has both broad and narrow emission lines above a strong blue
continuum, plus emission from the stellar component of the underlying
elliptical galaxy (Danziger \etal\ 1979), and so has some resemblance
to a low-power, flat radio-spectrum, quasar, although these emission
lines are relatively weak and the overall appearance of the spectrum
is very similar to that of the well-known BL~Lac 3C~371 (Angel \&
Stockman 1980), in which \chandra\ has also mapped jet X-ray
emission (Pesce \etal\ 2001). It is therefore important that more
low-redshift BL~Lac objects selected to cover a range of properties,
be well-imaged by \chandra\ to see whether similar X-ray jets and
halos can be found.

\section*{Acknowledgments}

MB and DMW are grateful to the staff of the ATCA for their assistance
during the observing run. The Australia Telescope is funded by the
Commonwealth of Australia for operation as a National Facility,
managed by CSIRO.


\label{lastpage}

\end{document}